\documentclass[aps,prl,twocolumn,noshowpacs,superscriptaddress,longbibliography]{revtex4}
\usepackage{CJK}
\usepackage{bm}
\usepackage{graphicx}
\usepackage{color}
\usepackage{amsmath,amssymb,amsfonts,amsthm}
\usepackage[colorlinks=true,linkcolor=blue,anchorcolor=red,citecolor=blue,urlcolor=blue]{hyperref}

\begin{document}

\title{Observation of anomalous $\pi$ modes in photonic Floquet engineering}

\author{Qingqing Cheng}
\affiliation{Shanghai Key Lab of Modern Optical System and Engineering Research Center of Optical Instrument and System (Ministry of Education), University of Shanghai for Science and Technology, Shanghai 200093, China}

\author{Yiming Pan}
\email{ yimingpan@mail.tau.ac.il}
\affiliation{Department of Electrical Engineering Physical Electronics, Tel Aviv University, Ramat Aviv 69978, Israel}

\author{Huaiqiang Wang}
\affiliation{National Laboratory of Solid State Microstructures and School of Physics, Nanjing University, Nanjing 210093, China}

\author{Chaoshi Zhang}
\affiliation{Shanghai Key Lab of Modern Optical System and Engineering Research Center of Optical Instrument and System (Ministry of Education), University of Shanghai for Science and Technology, Shanghai 200093, China}

\author{Dong Yu}
\affiliation{Shanghai Key Lab of Modern Optical System and Engineering Research Center of Optical Instrument and System (Ministry of Education), University of Shanghai for Science and Technology, Shanghai 200093, China}

\author{Avi Gover}
\affiliation{Department of Electrical Engineering Physical Electronics, Tel Aviv University, Ramat Aviv 69978, Israel}

\author{Haijun Zhang}
\affiliation{National Laboratory of Solid State Microstructures and School of Physics, Nanjing University, Nanjing 210093, China}

\author{Tao Li}
\affiliation{National Laboratory of Solid State Microstructures, School of Physics and Collaborative Innovation Centre of Advanced Microstructure, Nanjing University, Nanjing 210093, China}

\author{Lei Zhou}
\affiliation{State Key Laboratory of Surface Physics, Key Laboratory of Micro and Nano Photonic Structures (Ministry of Education) and Physics Department, Fudan University, Shanghai 200433, China}

\author{Shining Zhu}
\email{zhusn@nju.edu.cn}
\affiliation{National Laboratory of Solid State Microstructures, School of Physics, College of Engineering and Applied Sciences, Nanjing University, Nanjing 210093, China}
\date{\today }

\begin{abstract}
Recent progresses on Floquet topological phases have shed new light on time-dependant quantum systems, among which one-dimensional (1D) Floquet systems have been under extensive theoretical research. However, an unambiguous experimental observation of these 1D Floquet topological phases has still been lacking. Here, by periodically bending ultrathin metallic arrays of coupled corrugated waveguides, a photonic Floquet simulator was well designed and successfully fabricated to simulate the periodically driven Su-Schrieffer-Heeger model. Intriguingly, under moderate driven frequencies, we first experimentally observed and theoretically verified the Floquet topological $\pi$ mode propagating along the array's boundary. The different evolutionary behaviors between static and non-static topological end modes have also been clearly demonstrated. Our experiment also reveals the universal high-frequency behavior in perically driven systems. We emphasize that, our system can serve as a powerful and versatile testing ground for various phenomena related to time-dependant 1D quantum phases, such as Thouless charge pumping and manybody localization.
\end{abstract}
\maketitle

\emph{Introduction.}
Recently, following the development of topological insulators \cite{Hasan2010, Qi2010}, it has been shown that periodic perturbations (or modulations) can be used to realize new engineered topologically nontrivial phases not accessible in static equilibrium systems. This new area of research, termed ``Floquet engineering'' \cite{Andr2015, Novi2017, Marin2015, Phillip2017, Goldman2014, Khemani2016, oka2018floquet, Rodriguez2018}, has motivated growing interest in periodically driven quantum systems, which have been employed (naturally or artificially) in ultra-cold atom optical lattices \cite{Andr2015, Marin2015}, spin systems \cite{Marin2015}, time crystals~\cite{Shapere2012, Wilczek2012, Else2016, Bomantara2018} and photonic simulations \cite{Garanovich2012, Longhi2008, Longhi2005, Longhi2003, Cheng2014, Martin2016}. Intriguingly, recent theoretical studies have demonstrated that the Floquet quasi-energy spectrum of periodically driven systems feature richer topological structures and invariants than their non-driven counterparts~\cite{Kitagawa2010, Roy2016, Rudner2013, Potter2016, Else2016Class, Asb2014, DalLago2015, Platero2013, Fruchart2016, Frederik2015, Roy2017, Iadecola2015, Carpentier2015, Ho2014, Bomantara2016}, attaching to the gaps of quasi-energy bands, such as the Floquet-Majorana end state~\cite{Jiang, Thakurathi2013, Kundu2013, Wang2017}, topologically nontrivial zero or $\pi$ mode \cite{Asb2014, DalLago2015, Fruchart2016}, and topological singularities~\cite{Frederik2015}. However, finding materials that can be used to experimentally realize such Floquet-engineered topological phases remains a serious challenge in condensed matter physics~\cite{Andr2015, Marin2015, Wang2013Observation, Lindner2010Floquet}.

Inspired by great successes in the discoveries of topological phenomena in artificial quantum systems, such as ultra-cold atoms and photonic simulations~\cite{Kraus2012, Khanikaev2013Photonic, khanikaev2017two, Lu2014Topological}, many researchers have begun to explore experimental possibilities of achieving non-static engineered topological states in quantum systems. For example, two-dimensional (2D) photonic and phononic Floquet topological phases (FTP) have already been realized~\cite{Rechtsman2013Photonic, maczewsky2017observation}, with clear observations of Floquet topological edge states. The photonic lattice provides a window into Floquet topological physics, since its structure can be designed at will, which is not subject to structural defects or absorbate contamination. However, experimental realizations of the seemingly much simpler one-dimensional (1D) FTP have still been lacking, not to mention an unambiguous demonstration of corresponding 1D Floquet end modes (FEM). Moreover, the realistic dynamic evolution of the FEM has remained unclear so far, though it is supposed to be different from its static counterpart.

To address these problems, in this letter, we designed and fabricated a photonic Floquet simulator (PFS) to mimic a typical 1D Floquet system, namely, the periodically driven Su-Schrieffer-Heeger (SSH) model~\cite{Asb2014, Fruchart2016, DalLago2015}, through periodically bending ultrathin metallic arrays of coupled corrugated waveguides, which support spoof surface plasmon polaritons (SSPPs) at microwave to infrared wavelength. By tuning the bending periods and fixing initial input positions, we experimentally observed and theoretically verified the Floquet topological $\pi$ mode, which propagates along the array's boundary. We also give a clear demonstration of the different evolutionary behaviors between static and non-static topological end modes. In addition, our system reveals the universal high-frequency behavior prevailing in Floquet engieering~\cite{Andr2015}.  It should be emphasized that our system provides a versatile platform for investigating various phenomena related to time-dependant 1D quantum phases, such as Thouless charge pumping and manybody localization.

\begin{figure}
  \centering
  \includegraphics[width=3.3in]{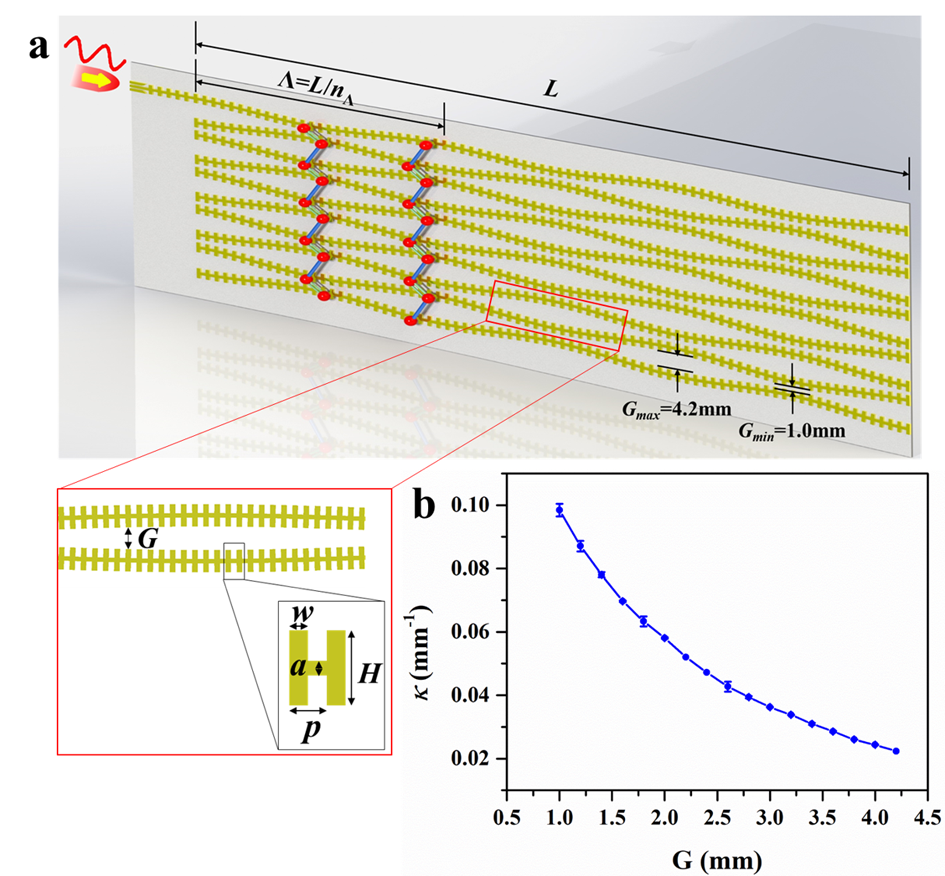}\\
  \caption{ (a) Schematic illustration of the experimental configuration for the periodically-driven SSH model based on well-designed periodically bending ultrathin metallic waveguides, with a cosine modulation of the spacing $G$ between two adjacent waveguides in the propagation direction $z$ as $G(z)=g_{0}\pm2A_{0}\cos(\frac{2\pi z}{\Lambda}+\theta_{0})$. The parameters are chosen as $L=400$mm, $g_{0}\approx2.6$mm, and $A_{0}\approx0.8$mm, yielding $G_{\mathrm{max}}\approx4.2$mm and $G_{\mathrm{min}}\approx1$mm. Inset: Illustration of the ``H''-shape structure, with the parameters given by $P=2$ mm, $H=4$ mm, $a=0.8$ mm, and $w=1$ mm. (b) The effective coupling constant $\kappa$ as a function of $G$. }\label{Fig1}
\end{figure}

{\emph{Periodically driven SSH model.}} The original SSH model~\cite{Su1979} is a well-known 1D topological structure with two degenerate ground-states differing by the relative strength between intra-cell and inter-cell hoppings. To study the periodically driven SSH model, we periodically bend the waveguide array along its propagating direction $z$,  as shown in Fig. 1(a), with opposite axis offsets $x_{0}(z)=\pm A_{0}\cos(\frac{2\pi z}{\Lambda}+\theta_{0})$ for two neighboring waveguides, where $A_{0}$ ($=0.8$ mm) and $\Lambda$ denote the amplitude and period of the sinusoidal bending, respectively, and $\theta_{0}$ is the initial phase which depends on the starting point $z_0$ during one period. Through the coupled-mode theory~\cite{Garanovich2012, Cheng2014} (see the supplemental material (SM)), the waveguide array can be accurately mapped into an effective $1+1$ dimensional tight-binding-approximated Hamiltonian as
\begin{equation}
\label{H}
H(z)=\sum_{i=1}^{N}\beta_{i}(z)c_{i}^{\dagger}c_{i}+\sum_{i=1}^{N-1}
\Big(\kappa_{0}+(-1)^{i}\Delta\kappa(z)\Big)c_{i}^{\dagger}c_{i+1}+\mathrm{H.c},
\end{equation}
where the guiding propagation $z$ acts as the time axis $t$. Here, $N$ is the number of waveguides, $\beta_{i}(z)$ is the effective propagation constant which can be reasonably treated as a constant $\beta_{0}$ in the weak-guidance approximation (WGA) of our configuration~\cite{Garanovich2012, Cheng2014}.  The second term in Eq. (\ref{H}) represents couplings between nearest-neighbor (NN) waveguides with a constant (staggered) coupling strength $\kappa_{0}$ ($\Delta \kappa(z)$). According to the WGA, the NN coupling strength $\kappa$ mainly depends on their distance $G(z)$, with the relation shown in Fig. 1 (b). Since $G(z)=g_{0}\pm2A_{0}\cos(\frac{2\pi z}{\Lambda}+\theta_{0})$ in our configuration, where $g_{0}$ ($\approx2.6$ mm) is the initial NN spacing without bending, the NN coupling strength can be approximated as $\kappa_{0}\pm\delta\kappa\cos(\frac{2\pi z}{\lambda}+\theta_{0})$, with the optimal parameters given by $\kappa_{0}\approx0.042\ \mathrm{mm}^{-1}$  and $\delta\kappa\approx0.02\ \mathrm{mm}^{-1}$. Consequently, $H(z)$ in Eq. (\ref{H}) exactly simulates a Floquet SSH model with time-periodic staggered NN coulpings.

\emph{Experimental Results.} In our experiments, we choose $N=10$ waveguides with their length set as $L=400$ mm ($L=800$ mm in the SM), and the driven frequency is given by $\omega\equiv\frac{2\pi}{\Lambda}=n_{\Lambda}\omega_{L}$, where $n$ is the total number of periods within the length $L$, and $\omega_{L}\equiv\frac{2\pi}{L}$ is the characteristic frequency. Thus, $n_{\Lambda}$ will be termed as the reduced frequency, which can be easily tuned to investigate various frequency-dependant phenomena as well as topological edge modes related to the Floquet SSH model~\cite{Asb2014, Fruchart2016, DalLago2015}, as we show in detail below. In addition, unless stated explicitly, the initial phase $\theta$ is set as zero.

\begin{figure*}
  \centering
  \includegraphics[width=6in]{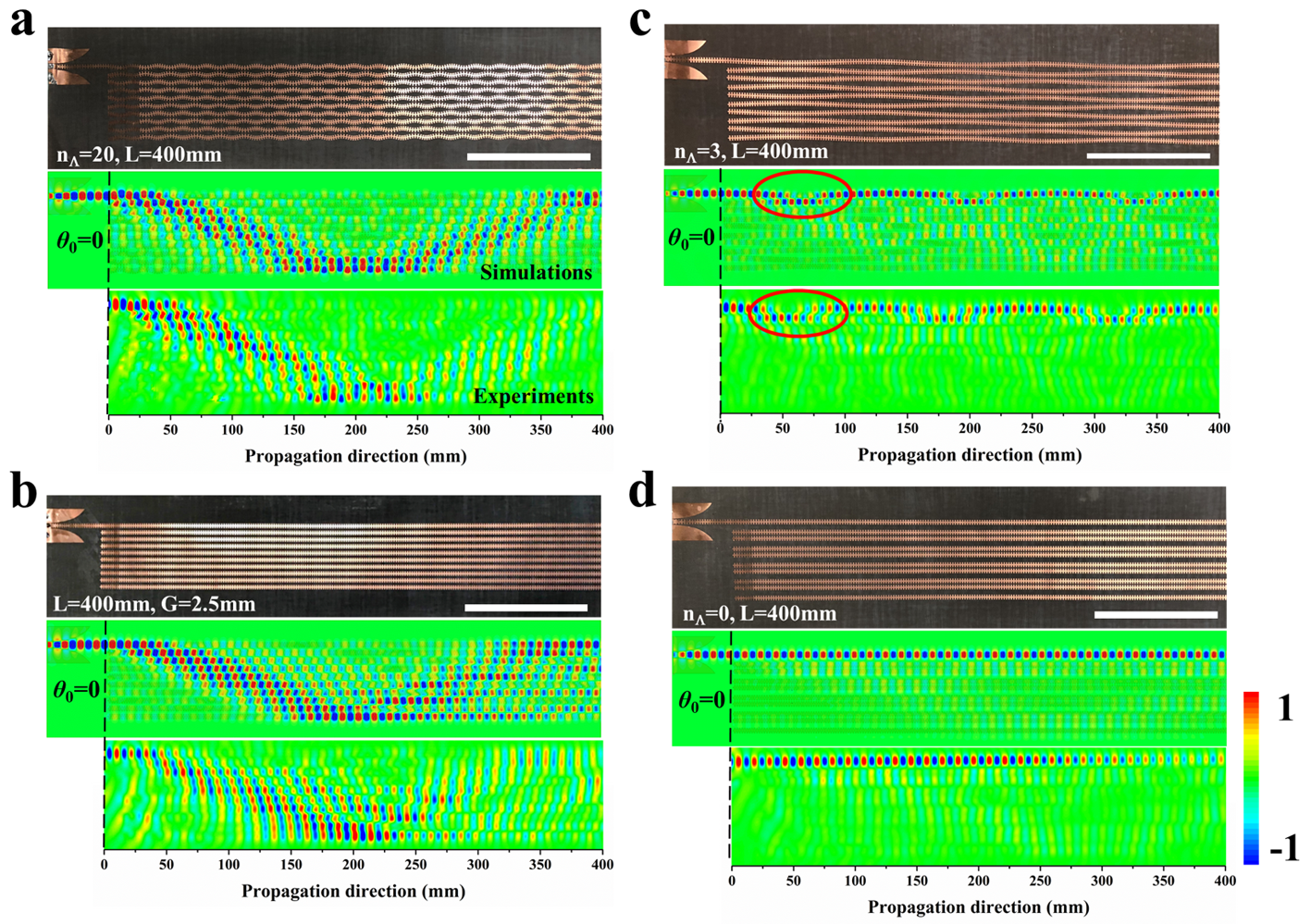}\\
  \caption{ (From up to down in each figure) Snapshots of the fabricated samples, CST simulation results, and near-field measurement of the intensity profile after injecting light from the upmost boundary waveguide, under different driving conditions with the same length $L=400$mm, waveguide number $N=10$ and initial phase $\theta_{0}=0$. Results for curved waveguides with (a) $n_{\Lambda}=20$ (topologically trivial high-frequency case) and (c) $n_{\Lambda}=3$ (topologically nontrivial case with anomalous boundary $\pi$-modes), and for straight waveguides ($n_{\Lambda}=0$) with (b) identical spacings $G=2.5$mm (trivial static case) and (d) dimerized spacings (nontrivial static case with boundary zero-modes) between nearest-neighboring waveguides.}\label{Fig2}
\end{figure*}

We start from the high frequency case where the driving frequency is much smaller than the effective coupling length and the time-periodic staggered NN coupling should be smeared out due to its fast oscillating behavior, thus rendering the system similar to that composed of straight waveguides with identical NN couplings, which obviously belongs to the trivial phase.  A waveguide array with a representative $\omega=20\omega_{L}$ ($\approx0.306\ \mathrm{mm}^{-1}$ ) in this high-frequency range was fabricated, as illustrated in Fig. 2a, to carry out the near-field measurement of the intensity profile after injecting light from the upmost boundary waveguide. This propagation pattern resembles that of straight waveguides with identical NN couplings, as presented in Fig. 2b for comparison. However, the micro-pattern of the $\omega=20\omega_{L}$ case is different from that of the undimerized straight waveguides with smooth diffraction management.  This stems from the micro-motion due to the finite-frequency effect in the fast-driving regime.

Intriguingly, as we gradually lower the driving frequency to the range $\omega=2\sim5\omega_{L}$ ($0.0306\sim0.0766\ \mathrm{ mm}^{-1}$), a quite distinct propagation pattern arises, as exemplified by the $\omega=3\omega_{L}$ ($0.0460\ \mathrm{ mm}^{-1}$) structure with corresponding experimental results in Fig. 2c (see the SM for other frequencies). The injected light no longer spreads into the bulk array, but instead is mainly localized within the two waveguides at the up boundary. Moreover, the intensity profile exhibits a periodic oscillation pattern in its distribution between the two boundary guides. This periodically oscillating anomalous edge mode is the main finding of our work, which will later be theoretically proved as the long pursued Floquet $\pi$-mode predicted in the Floquet SSH model~\cite{Asb2014, Fruchart2016, DalLago2015}. To further reveal its difference from the well-known zero-mode edge state of the static SSH model, we fabricated an array of straight waveguides with time-independent staggered NN couplings (see Fig. 2d), which lies in the topological nontrivial phase~\cite{Su1979, Cheng2014}. The experimental measurements in Fig. 2d shows that the injected light always propagates along the boundary waveguide without spreading into the bulk or displaying any oscillation in its intensity distribution, in striking contrast to the above nonstatic edge mode propagation. The non-spreading stroboscopic evolution for the Floquet $\pi$-mode is the first time that a 1D periodically driven end state has been realized and observed experimentally in the photonic platform.

\begin{figure*}
  \centering
  \includegraphics[width=6.3in]{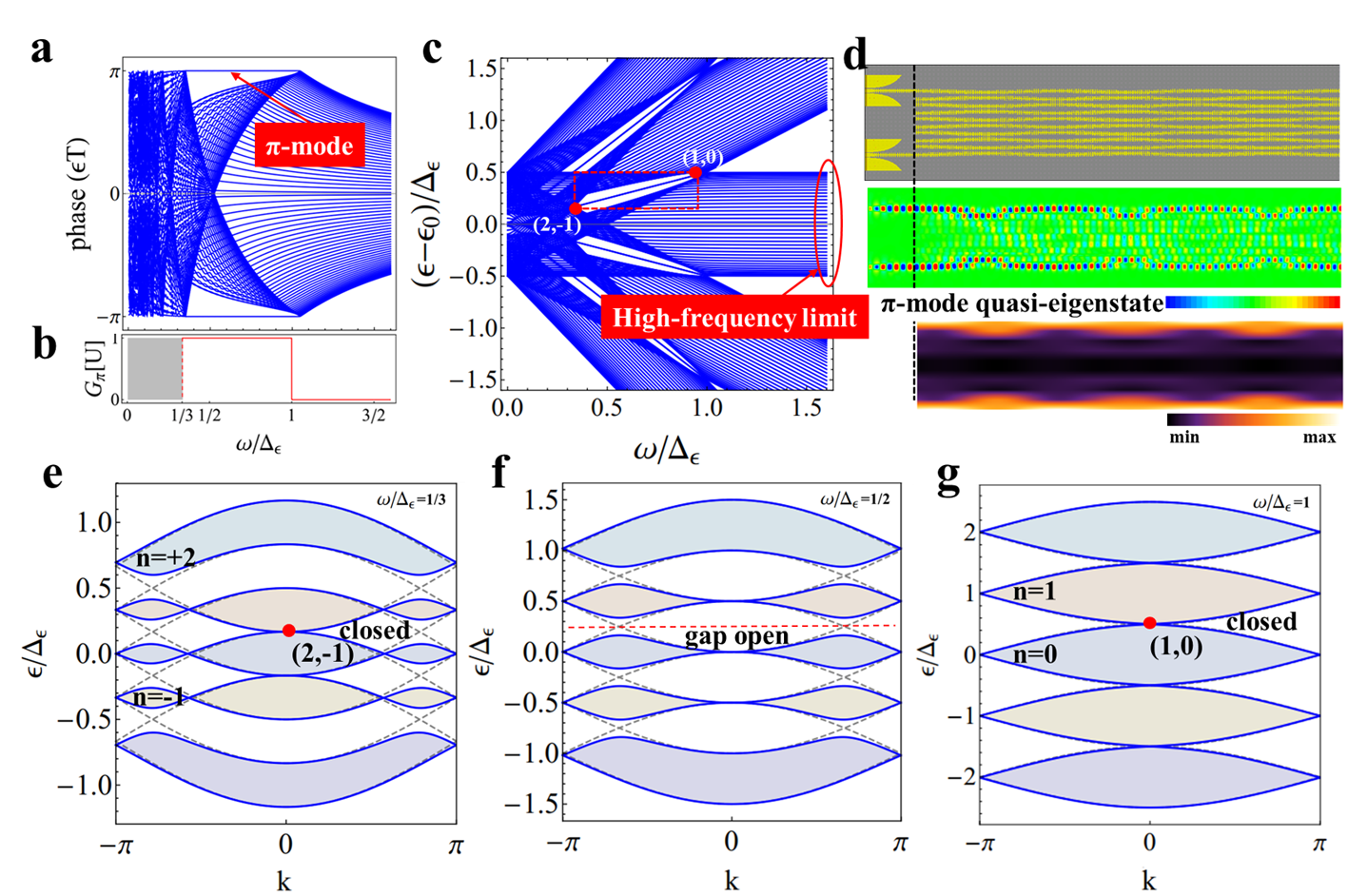}\\
  \caption{(a) Quasienergy under OBC with 40 waveguides and (b) $G_{\pi}$ as a function of the driving frequency $\omega$, where the bandwidth $\Delta$ is taken as the energy unit. Floquet $\pi$-modes can be found in (a) when $1/3<\omega/\Delta<1$, $G_{\pi}=1$, as confirmed by the value of $G_{\pi}$ in (b). By choosing five frequency replicas $n=0,\pm1,\pm2$ in the frequency space, (c) quasienergy under OBC with 40 waveguides as a function of $\omega/\Delta$, (d) dynamic evolution of the $\pi$-mode state for 10 waveguides and $n_{\Lambda}=3$, where the waveguides' configuration and CST simulation result are presented for reference. The quasienergy band structures of the five chosen frequency replicas with (e) $\omega/\Delta=1/3$, (f) $\omega/\Delta=1/2$, and (g) $\omega/\Delta=1$.}\label{Fig3}
\end{figure*}

\emph{Theoretical analysis.} To verify the existence of the anomalous edge mode, we explicitly calculate the topological invariant of our system to determine its phase diagram by using Floquet theory. The time evolution operator of the system is given as $U(z,z_{0})=\hat{T}e^{-i\int_{z_{0}}^{z}H(z')dz'}$,  where $\hat{T}$ denotes the time-ordering operator, and $z_{0}$ is the starting point. Without loss of generality, we set $z_{0}=0$ and $U(z,z_{0})=U(z)$ to simplify notations. The Floquet operator is defined as the time evolution operator for one full period, given by $U(\Lambda)$~\cite{Lindner2010Floquet}, from which a time-averaged effective Hamiltonian can be defined as $H_{\mathrm{eff}}=\frac{i}{\Lambda}\ln U(\Lambda)$~\cite{Lindner2010Floquet}. The eigenvalues of $H_{\mathrm{eff}}$ corresponds to the quasienergy spectrum of the system. Due to the translation symmetry, both $U(z)$ and $H_{\mathrm{eff}}$ can be Bloch decomposed as $U_{z}=\prod U(z,k)$ and $H_{\mathrm{eff}}=\sum_{k}H_{\mathrm{eff}}(k)$, respectively. According to Ref. \cite{Fruchart2016}, a $\mathbb{Z}$-valued invariant can be defined for the quasienergy gap at zero or $\pi$ for a 1D periodically driven system with chiral symmetry, as is also satisfied in our system. To calculate this invariant, we must resort to the periodized evolution operator, given by $V(z,k)\equiv U(z,k)e^{iH_{\mathrm{eff}}(k)z}$. Since the zero quasienergy gap is found to be always closed in our system, regardless of the value of $\omega$, we only need to calculate the $\pi$ gap invariant $G_{\pi}$ through the following formula~\cite{Asb2014, Fruchart2016, DalLago2015}:
\begin{equation}
\label{Gpi}
G_{\pi}=\frac{i}{2\pi}\int_{-\pi}^{\pi}\mathrm{tr}\Big((V^{+}_{\pi})^{-1}\partial_{k} V^{+}_{\pi}\Big)dk,
\end{equation}
where $V^{+}_{\pi}$ is obtained from $V(z,k)$ at half period:
\begin{equation}
\label{V}
V(\Lambda/2,k)=\left(
                         \begin{array}{cc}
                           V_{\pi}^{+} & 0 \\
                           0 & V_{\pi}^{-} \\
                         \end{array}
                       \right).
\end{equation}
The numerical result of $G_{\pi}$ as a function of $\omega$ is presented in Fig. 3b, where the bandwidth of the undriven system $\Delta=4\kappa_{0}$ is taken as the energy unit for reasons that will be made clear later. When $\omega/\Delta>1$, $G_{\pi}=0$, while for $1/3<\omega/\Delta<1$, $G_{\pi}=1$. (our numerical result of $G_{\pi}$ for $\omega/\Delta<1/3$ is not shown here since it no longer takes desired integer values, which may be related to the complex low-frequency behavior~\cite{Rodriguez2018}.) A nonzero $G_{\pi}$ indicates a topologically nontrivial phase and through bulk-edge correspondence, it corresponds to the number of edge $\pi$ mode within the $\pi$ gap~\cite{Fruchart2016}. This is confirmed by the open-boundary quasienergy spectrum shown in Fig. 3a for $N=40$ waveguides, where $\pi$ modes exist for $1/3<\omega/\Delta<1$. It should be noted that the experimental range of $\omega=0.0306\sim0.0766\ \mathrm{ mm}^{-1}$, where anomalous edge modes are observed, approximately falls into this topologically nontrivial region of $\Delta/3\sim\Delta$ ($0.042\sim0.168\ \mathrm{ mm}^{-1}$), thus permiting us to reasonably identify these anomalous modes as the $\pi$ mode predicted in the periodically driven SSH model~\cite{Frederik2015}.

In fact, the underlying physics becomes much clearer in the direct-product Floquet space: $\mathcal{H}\otimes\mathcal{T}$~\cite{Platero2013, DalLago2015}, where $\mathcal{H}$ is the usual Hilbert space and $\mathcal{T}$ denotes the space of time-periodic functions spanned by $e^{in\omega t}$, with the integer index $n$ representing the $n$-th Floquet replica~\cite{DalLago2015}. Considering the periodic nature of the energy, we only need to focus on the energy range $(-\pi,\pi]$ (or $(-\frac{\omega}{2},\frac{\omega}{2}]$) of the $n=0$ replica. When $\omega>\Delta$, $n=0$ replica is decoupled from other replicas with a gap between the $n=1$ replica at $\epsilon=\pi$ ($-\pi$ is equivalent to $\pi$), which should be topologically trivial since this gap persists when $\omega$ approaches the trivial high-frequency limit. At $\omega=\Delta$, this gap is closed due to the band touching of the $n=0$ and $n=1$ replicas at $\pi$, as shown in Fig. 3g. With further decreasing $\omega$, the coupling between $n=1$ and $n=0$ replicas opens this gap again, as shown in Fig. 3f with $\omega=\Delta/2$, and it does not close until $\omega=\Delta/3$, where the $n=2$ and $n=-1$ replicas touch each other at $\pi$ (see Fig. 3e). It has been shown in Ref.~\cite{DalLago2015} that this gap closing and reopening process switches the $\pi$ gap from trivial to nontrivial through the calculation of the Zak phase~\cite{DalLago2015}. Consequently, the same nontrivial region of $\Delta/3<\omega<\Delta$ is obtained as above from $G_{\pi}$. When $\omega$ is smaller than $\Delta/3$, more and more Floquet replicas should be involved~\cite{DalLago2015}, which is beyond the scope of this paper. It is worth mentioning that the zero quasienergy gap is always closed regardless of $\omega$, thus eliminating the possibility of the emergence of in-gap zero quasienergy modes. To further support the above argument, in Fig. 3c, we choose $N=40$ waveguides, $\omega=3\omega_{L}$, and $n=0,\pm1,\pm2$ replicas to numerically plot the open-boundary spectrum and in Fig. 3d we choose 10 waveguides to plot the time-dependant evolution of the $\pi$ edge modes. It is obvious that these modes propagate along the array's boundaries and exhibit periodic oscillation in the intensity distribution, which matches very well with the CST simulation results presented above for reference.

\begin{figure}
  \centering
  \includegraphics[width=3.3in]{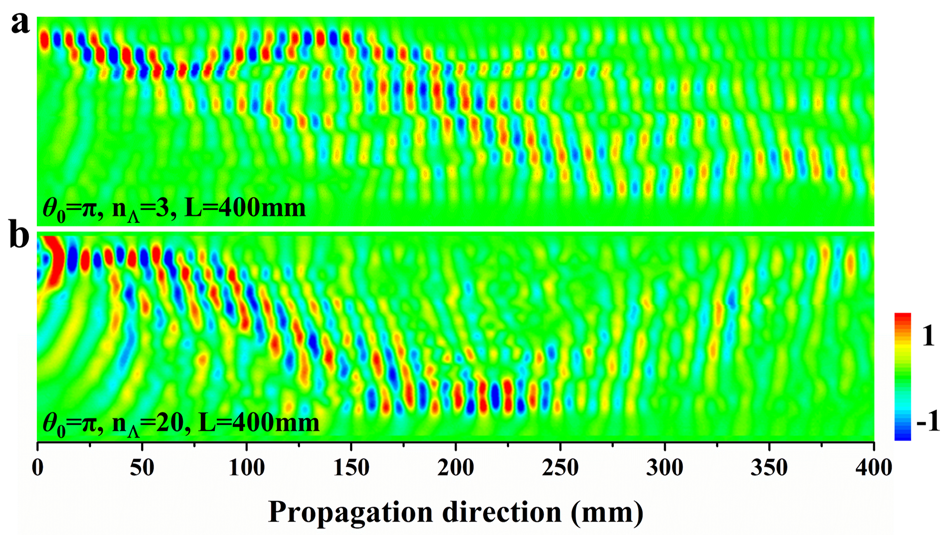}\\
  \caption{Near-field measurements of the intensity profile for 10 waveguide of length $L=400$mm, initial phase $\theta_{0}=\pi$, and (a) $n_{\Lambda}=3$ and (b) $n_{\Lambda}=20$. Compared to Fig. 2(c) with $\theta_{0}=0$, the injected light in (a) no longer propagates along the boundary waveguide, thus indicating the absence of the $\pi$-mode. However, the propagation pattern in (b) is very similar to that in Fig. 2a, which is related to the universal high-frequency behavior.}\label{Fig4}
\end{figure}

\emph{Discussion.} First, we discuss the dependence of the initial phase $\theta_{0}$, which can be easily tuned by simply adjusting the initial input position in our experiment. It is found that, although the quasienergy spectrum as well as the existence of the $\pi$-mode is theoretically independent of $\theta_0$, to experimentally excite and observe the photonic $\pi$-mode, $\theta_0$ must be tuned to the region $[-\pi/2,\pi/2]$, i.e., the instantaneous Hamiltonian at $z=0$ must lie in the topological nontrivial phase. This is illustrated by our experimental measurements in Figs. 2c and 4a for $\theta_{0}=0$ and $\pi$, respectively, under the same driving frequency $\omega=3\omega_{L}$ (results for other $\theta_{0}$ values are presented in the SM). In contrast to Fig. 2c, in Fig. 4a, the injected light no longer propagates along the boundary waveguide, but gradually spreads into the bulk array, thus indicating no excitation of the $\pi$-mode. In addition, for the high-frequency case, where the system lies in the topologically trivial phase, the propagating pattern shows no dependence of $\theta_{0}$, as can be seen by comparing Fig. 2a ($\theta_{0}=0$) and Fig. 4b ($\theta_{0}=\pi$), with the same $\omega=20\omega_{L}$. This is related to the universal high-frequency behavior of a periodically driven system~\cite{Andr2015}.

Second, we examine the topological robustness of the $\pi$-mode in the presence of weak disorder by introducing weak random coupling coefficients between neighboring waveguides $\delta\kappa=\kappa_{\mathrm{min}}=0.042\mathrm{mm} ^{-1}$. Through the CST simulation, in Figs. 5 (a) and 5(b), we present the light propagation pattern for $\theta_{0}=0$ and $\omega=3\omega_{L}$ and $\omega=4\omega_{L}$, respectively. It is obvious that the $\pi$-mode still propagates along the boundary with periodic oscillation and negligible amount of dissipation into the bulk, which indeed suggests its robustness to weak disorder.

\begin{figure}
  \centering
  \includegraphics[width=3in]{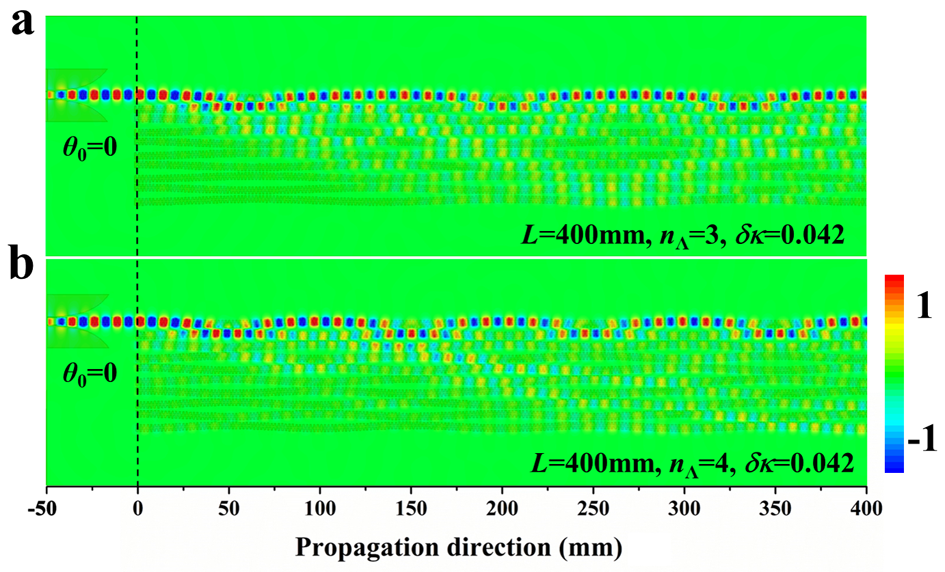}\\
  \caption{CST simulation results for the weakly disordered samples with $L=400$mm, $\theta_{0}=0$, and (a) $n_{\Lambda}=3$ and (b) $n_{\Lambda}=4$. Here, disorder effects are introduced by imposing weak random coupling coefficient $\delta\kappa=\kappa_{\mathrm{min}}=0.042\mathrm{mm} ^{-1}$. Both (a) and (b) demonstrate the topological robustness of the $\pi$-mode edge state under weak disorder.}\label{Fig5}
\end{figure}

\emph{Conclusion.} In summary, by periodically bending ultrathin metallic arrays of coupled corrugated waveguides, we have successfully realized a photonic simulation of the 1D periodically SSH model. Under certain frequency range and initial input positions, we first experimentally observed the long-pursued Floquet $\pi$-mode in 1D periodically driven systems, which has been further verified and well explained by the Floquet theory. Due to the high flexibility and tunability of our system, it can be used to investigate various phenomena in 1D time-dependent quantum systems, including both the adiabatic Thouless charge pump~\cite{Rice, Shen2017Topological, Zhou2015, Wang2015Interband} and the possible nonadiabatic Floquet charge pump~\cite{Ho2012}, which will be our future work.

\emph{Acknowledgement}. The authors thank Che-Ting Chan, Ady Arie, Jiangbin Gong, and Koby Scheuer for their helpful discussions and comments. In particular, we thank Xiaopeng Shen for helping us measure the intensity distributions and discuss experiments. This work was supported by the National Natural Science Foundation of China (Grant Nos. 11604208, 11674167, 11621091, 11322439, 11734007, 11474057, 11674165 and 11674068), Shanghai Science and Technology Committee (Grant Nos. 16ZR1445600 and 16JC1403100), National Key Research and Development Program of China (Grant Nos. 2017YFA0303700 and 2017YFA0303504), and in part by DIP (German-Israeli Project Cooperation) and the US-Israel Binational Science Foundation and by the PBC programme of the Israel Council of Higher Education.
\bibliography{reference}

\begin{thebibliography}{53}
\expandafter\ifx\csname natexlab\endcsname\relax\def\natexlab#1{#1}\fi
\expandafter\ifx\csname bibnamefont\endcsname\relax
  \def\bibnamefont#1{#1}\fi
\expandafter\ifx\csname bibfnamefont\endcsname\relax
  \def\bibfnamefont#1{#1}\fi
\expandafter\ifx\csname citenamefont\endcsname\relax
  \def\citenamefont#1{#1}\fi
\expandafter\ifx\csname url\endcsname\relax
  \def\url#1{\texttt{#1}}\fi
\expandafter\ifx\csname urlprefix\endcsname\relax\def\urlprefix{URL }\fi
\providecommand{\bibinfo}[2]{#2}
\providecommand{\eprint}[2][]{\url{#2}}

\bibitem[{\citenamefont{Hasan and Kane}(2010)}]{Hasan2010}
\bibinfo{author}{\bibfnamefont{M.~Z.} \bibnamefont{Hasan}} \bibnamefont{and}
  \bibinfo{author}{\bibfnamefont{C.~L.} \bibnamefont{Kane}},
  \bibinfo{journal}{Rev. Mod. Phys.} \textbf{\bibinfo{volume}{82}},
  \bibinfo{pages}{3045} (\bibinfo{year}{2010}).

\bibitem[{\citenamefont{Qi and Zhang}(2011)}]{Qi2010}
\bibinfo{author}{\bibfnamefont{X.-L.} \bibnamefont{Qi}} \bibnamefont{and}
  \bibinfo{author}{\bibfnamefont{S.-C.} \bibnamefont{Zhang}},
  \bibinfo{journal}{Rev. Mod. Phys.} \textbf{\bibinfo{volume}{83}},
  \bibinfo{pages}{1057} (\bibinfo{year}{2011}).

\bibitem[{\citenamefont{Eckardt and Anisimovas}(2015)}]{Andr2015}
\bibinfo{author}{\bibfnamefont{A.}~\bibnamefont{Eckardt}} \bibnamefont{and}
  \bibinfo{author}{\bibfnamefont{E.}~\bibnamefont{Anisimovas}},
  \bibinfo{journal}{New Journal of Physics} \textbf{\bibinfo{volume}{17}},
  \bibinfo{pages}{093039} (\bibinfo{year}{2015}).

\bibitem[{\citenamefont{Novi\ifmmode~\check{c}\else \v{c}\fi{}enko
  et~al.}(2017)\citenamefont{Novi\ifmmode~\check{c}\else \v{c}\fi{}enko,
  Anisimovas, and Juzeli\ifmmode~\bar{u}\else \={u}\fi{}nas}}]{Novi2017}
\bibinfo{author}{\bibfnamefont{V.}~\bibnamefont{Novi\ifmmode~\check{c}\else
  \v{c}\fi{}enko}},
  \bibinfo{author}{\bibfnamefont{E.}~\bibnamefont{Anisimovas}},
  \bibnamefont{and}
  \bibinfo{author}{\bibfnamefont{G.}~\bibnamefont{Juzeli\ifmmode~\bar{u}\else
  \={u}\fi{}nas}}, \bibinfo{journal}{Phys. Rev. A}
  \textbf{\bibinfo{volume}{95}}, \bibinfo{pages}{023615}
  (\bibinfo{year}{2017}).

\bibitem[{\citenamefont{Bukov et~al.}(2015)\citenamefont{Bukov, D'Alessio, and
  Polkovnikov}}]{Marin2015}
\bibinfo{author}{\bibfnamefont{M.}~\bibnamefont{Bukov}},
  \bibinfo{author}{\bibfnamefont{L.}~\bibnamefont{D'Alessio}},
  \bibnamefont{and}
  \bibinfo{author}{\bibfnamefont{A.}~\bibnamefont{Polkovnikov}},
  \bibinfo{journal}{Advances in Physics} \textbf{\bibinfo{volume}{64}},
  \bibinfo{pages}{139} (\bibinfo{year}{2015}).

\bibitem[{\citenamefont{Weinberg et~al.}(2017)\citenamefont{Weinberg, Bukov,
  D¡¯Alessio, Polkovnikov, Vajna, and Kolodrubetz}}]{Phillip2017}
\bibinfo{author}{\bibfnamefont{P.}~\bibnamefont{Weinberg}},
  \bibinfo{author}{\bibfnamefont{M.}~\bibnamefont{Bukov}},
  \bibinfo{author}{\bibfnamefont{L.}~\bibnamefont{D¡¯Alessio}},
  \bibinfo{author}{\bibfnamefont{A.}~\bibnamefont{Polkovnikov}},
  \bibinfo{author}{\bibfnamefont{S.}~\bibnamefont{Vajna}}, \bibnamefont{and}
  \bibinfo{author}{\bibfnamefont{M.}~\bibnamefont{Kolodrubetz}},
  \bibinfo{journal}{Physics Reports} \textbf{\bibinfo{volume}{688}},
  \bibinfo{pages}{1 } (\bibinfo{year}{2017}).

\bibitem[{\citenamefont{Goldman and Dalibard}(2014)}]{Goldman2014}
\bibinfo{author}{\bibfnamefont{N.}~\bibnamefont{Goldman}} \bibnamefont{and}
  \bibinfo{author}{\bibfnamefont{J.}~\bibnamefont{Dalibard}},
  \bibinfo{journal}{Phys. Rev. X} \textbf{\bibinfo{volume}{4}},
  \bibinfo{pages}{031027} (\bibinfo{year}{2014}).

\bibitem[{\citenamefont{Khemani et~al.}(2016)\citenamefont{Khemani, Lazarides,
  Moessner, and Sondhi}}]{Khemani2016}
\bibinfo{author}{\bibfnamefont{V.}~\bibnamefont{Khemani}},
  \bibinfo{author}{\bibfnamefont{A.}~\bibnamefont{Lazarides}},
  \bibinfo{author}{\bibfnamefont{R.}~\bibnamefont{Moessner}}, \bibnamefont{and}
  \bibinfo{author}{\bibfnamefont{S.~L.} \bibnamefont{Sondhi}},
  \bibinfo{journal}{Phys. Rev. Lett.} \textbf{\bibinfo{volume}{116}},
  \bibinfo{pages}{250401} (\bibinfo{year}{2016}).

\bibitem[{\citenamefont{Oka and Kitamura}(2018)}]{oka2018floquet}
\bibinfo{author}{\bibfnamefont{T.}~\bibnamefont{Oka}} \bibnamefont{and}
  \bibinfo{author}{\bibfnamefont{S.}~\bibnamefont{Kitamura}},
  \bibinfo{journal}{arXiv:1804.03212}  (\bibinfo{year}{2018}).

\bibitem[{\citenamefont{Rodriguez-Vega and Seradjeh}(2018)}]{Rodriguez2018}
\bibinfo{author}{\bibfnamefont{M.}~\bibnamefont{Rodriguez-Vega}}
  \bibnamefont{and} \bibinfo{author}{\bibfnamefont{B.}~\bibnamefont{Seradjeh}},
  \bibinfo{journal}{Phys. Rev. Lett.} \textbf{\bibinfo{volume}{121}},
  \bibinfo{pages}{036402} (\bibinfo{year}{2018}).

\bibitem[{\citenamefont{Shapere and Wilczek}(2012)}]{Shapere2012}
\bibinfo{author}{\bibfnamefont{A.}~\bibnamefont{Shapere}} \bibnamefont{and}
  \bibinfo{author}{\bibfnamefont{F.}~\bibnamefont{Wilczek}},
  \bibinfo{journal}{Phys. Rev. Lett.} \textbf{\bibinfo{volume}{109}},
  \bibinfo{pages}{160402} (\bibinfo{year}{2012}).

\bibitem[{\citenamefont{Wilczek}(2012)}]{Wilczek2012}
\bibinfo{author}{\bibfnamefont{F.}~\bibnamefont{Wilczek}},
  \bibinfo{journal}{Phys. Rev. Lett.} \textbf{\bibinfo{volume}{109}},
  \bibinfo{pages}{160401} (\bibinfo{year}{2012}).

\bibitem[{\citenamefont{Else et~al.}(2016)\citenamefont{Else, Bauer, and
  Nayak}}]{Else2016}
\bibinfo{author}{\bibfnamefont{D.~V.} \bibnamefont{Else}},
  \bibinfo{author}{\bibfnamefont{B.}~\bibnamefont{Bauer}}, \bibnamefont{and}
  \bibinfo{author}{\bibfnamefont{C.}~\bibnamefont{Nayak}},
  \bibinfo{journal}{Phys. Rev. Lett.} \textbf{\bibinfo{volume}{117}},
  \bibinfo{pages}{090402} (\bibinfo{year}{2016}).

\bibitem[{\citenamefont{Bomantara and Gong}(2018)}]{Bomantara2018}
\bibinfo{author}{\bibfnamefont{R.~W.} \bibnamefont{Bomantara}}
  \bibnamefont{and} \bibinfo{author}{\bibfnamefont{J.}~\bibnamefont{Gong}},
  \bibinfo{journal}{Phys. Rev. Lett.} \textbf{\bibinfo{volume}{120}},
  \bibinfo{pages}{230405} (\bibinfo{year}{2018}).

\bibitem[{\citenamefont{Garanovich et~al.}(2012)\citenamefont{Garanovich,
  Longhi, Sukhorukov, and Kivshar}}]{Garanovich2012}
\bibinfo{author}{\bibfnamefont{I.~L.} \bibnamefont{Garanovich}},
  \bibinfo{author}{\bibfnamefont{S.}~\bibnamefont{Longhi}},
  \bibinfo{author}{\bibfnamefont{A.~A.} \bibnamefont{Sukhorukov}},
  \bibnamefont{and} \bibinfo{author}{\bibfnamefont{Y.~S.}
  \bibnamefont{Kivshar}}, \bibinfo{journal}{Physics Reports}
  \textbf{\bibinfo{volume}{518}}, \bibinfo{pages}{1 } (\bibinfo{year}{2012}).

\bibitem[{\citenamefont{Longhi and Staliunas}(2008)}]{Longhi2008}
\bibinfo{author}{\bibfnamefont{S.}~\bibnamefont{Longhi}} \bibnamefont{and}
  \bibinfo{author}{\bibfnamefont{K.}~\bibnamefont{Staliunas}},
  \bibinfo{journal}{Optics Communications} \textbf{\bibinfo{volume}{281}},
  \bibinfo{pages}{4343 } (\bibinfo{year}{2008}).

\bibitem[{\citenamefont{Longhi}(2005)}]{Longhi2005}
\bibinfo{author}{\bibfnamefont{S.}~\bibnamefont{Longhi}},
  \bibinfo{journal}{Opt. Lett.} \textbf{\bibinfo{volume}{30}},
  \bibinfo{pages}{2137} (\bibinfo{year}{2005}).

\bibitem[{\citenamefont{Longhi et~al.}(2003)\citenamefont{Longhi, Janner,
  Marano, and Laporta}}]{Longhi2003}
\bibinfo{author}{\bibfnamefont{S.}~\bibnamefont{Longhi}},
  \bibinfo{author}{\bibfnamefont{D.}~\bibnamefont{Janner}},
  \bibinfo{author}{\bibfnamefont{M.}~\bibnamefont{Marano}}, \bibnamefont{and}
  \bibinfo{author}{\bibfnamefont{P.}~\bibnamefont{Laporta}},
  \bibinfo{journal}{Phys. Rev. E} \textbf{\bibinfo{volume}{67}},
  \bibinfo{pages}{036601} (\bibinfo{year}{2003}).

\bibitem[{\citenamefont{Cheng et~al.}(2015)\citenamefont{Cheng, Pan, Wang, Li,
  and Zhu}}]{Cheng2014}
\bibinfo{author}{\bibfnamefont{Q.}~\bibnamefont{Cheng}},
  \bibinfo{author}{\bibfnamefont{Y.}~\bibnamefont{Pan}},
  \bibinfo{author}{\bibfnamefont{Q.}~\bibnamefont{Wang}},
  \bibinfo{author}{\bibfnamefont{T.}~\bibnamefont{Li}}, \bibnamefont{and}
  \bibinfo{author}{\bibfnamefont{S.}~\bibnamefont{Zhu}},
  \bibinfo{journal}{Laser \& Photonics Reviews} \textbf{\bibinfo{volume}{9}},
  \bibinfo{pages}{392} (\bibinfo{year}{2015}).

\bibitem[{\citenamefont{Holthaus}(2016)}]{Martin2016}
\bibinfo{author}{\bibfnamefont{M.}~\bibnamefont{Holthaus}},
  \bibinfo{journal}{Journal of Physics B: Atomic, Molecular and Optical
  Physics} \textbf{\bibinfo{volume}{49}}, \bibinfo{pages}{013001}
  (\bibinfo{year}{2016}).

\bibitem[{\citenamefont{Kitagawa et~al.}(2010)\citenamefont{Kitagawa, Berg,
  Rudner, and Demler}}]{Kitagawa2010}
\bibinfo{author}{\bibfnamefont{T.}~\bibnamefont{Kitagawa}},
  \bibinfo{author}{\bibfnamefont{E.}~\bibnamefont{Berg}},
  \bibinfo{author}{\bibfnamefont{M.}~\bibnamefont{Rudner}}, \bibnamefont{and}
  \bibinfo{author}{\bibfnamefont{E.}~\bibnamefont{Demler}},
  \bibinfo{journal}{Phys. Rev. B} \textbf{\bibinfo{volume}{82}},
  \bibinfo{pages}{235114} (\bibinfo{year}{2010}).

\bibitem[{\citenamefont{Roy and Harper}(2016)}]{Roy2016}
\bibinfo{author}{\bibfnamefont{R.}~\bibnamefont{Roy}} \bibnamefont{and}
  \bibinfo{author}{\bibfnamefont{F.}~\bibnamefont{Harper}},
  \bibinfo{journal}{Phys. Rev. B} \textbf{\bibinfo{volume}{94}},
  \bibinfo{pages}{125105} (\bibinfo{year}{2016}).

\bibitem[{\citenamefont{Rudner et~al.}(2013)\citenamefont{Rudner, Lindner,
  Berg, and Levin}}]{Rudner2013}
\bibinfo{author}{\bibfnamefont{M.~S.} \bibnamefont{Rudner}},
  \bibinfo{author}{\bibfnamefont{N.~H.} \bibnamefont{Lindner}},
  \bibinfo{author}{\bibfnamefont{E.}~\bibnamefont{Berg}}, \bibnamefont{and}
  \bibinfo{author}{\bibfnamefont{M.}~\bibnamefont{Levin}},
  \bibinfo{journal}{Phys. Rev. X} \textbf{\bibinfo{volume}{3}},
  \bibinfo{pages}{031005} (\bibinfo{year}{2013}).

\bibitem[{\citenamefont{Potter et~al.}(2016)\citenamefont{Potter, Morimoto, and
  Vishwanath}}]{Potter2016}
\bibinfo{author}{\bibfnamefont{A.~C.} \bibnamefont{Potter}},
  \bibinfo{author}{\bibfnamefont{T.}~\bibnamefont{Morimoto}}, \bibnamefont{and}
  \bibinfo{author}{\bibfnamefont{A.}~\bibnamefont{Vishwanath}},
  \bibinfo{journal}{Phys. Rev. X} \textbf{\bibinfo{volume}{6}},
  \bibinfo{pages}{041001} (\bibinfo{year}{2016}).

\bibitem[{\citenamefont{Else and Nayak}(2016)}]{Else2016Class}
\bibinfo{author}{\bibfnamefont{D.~V.} \bibnamefont{Else}} \bibnamefont{and}
  \bibinfo{author}{\bibfnamefont{C.}~\bibnamefont{Nayak}},
  \bibinfo{journal}{Phys. Rev. B} \textbf{\bibinfo{volume}{93}},
  \bibinfo{pages}{201103} (\bibinfo{year}{2016}).

\bibitem[{\citenamefont{Asb\'oth et~al.}(2014)\citenamefont{Asb\'oth,
  Tarasinski, and Delplace}}]{Asb2014}
\bibinfo{author}{\bibfnamefont{J.~K.} \bibnamefont{Asb\'oth}},
  \bibinfo{author}{\bibfnamefont{B.}~\bibnamefont{Tarasinski}},
  \bibnamefont{and} \bibinfo{author}{\bibfnamefont{P.}~\bibnamefont{Delplace}},
  \bibinfo{journal}{Phys. Rev. B} \textbf{\bibinfo{volume}{90}},
  \bibinfo{pages}{125143} (\bibinfo{year}{2014}).

\bibitem[{\citenamefont{Dal~Lago et~al.}(2015)\citenamefont{Dal~Lago, Atala,
  and Foa~Torres}}]{DalLago2015}
\bibinfo{author}{\bibfnamefont{V.}~\bibnamefont{Dal~Lago}},
  \bibinfo{author}{\bibfnamefont{M.}~\bibnamefont{Atala}}, \bibnamefont{and}
  \bibinfo{author}{\bibfnamefont{L.~E.~F.} \bibnamefont{Foa~Torres}},
  \bibinfo{journal}{Phys. Rev. A} \textbf{\bibinfo{volume}{92}},
  \bibinfo{pages}{023624} (\bibinfo{year}{2015}).

\bibitem[{\citenamefont{G\'omez-Le\'on and Platero}(2013)}]{Platero2013}
\bibinfo{author}{\bibfnamefont{A.}~\bibnamefont{G\'omez-Le\'on}}
  \bibnamefont{and} \bibinfo{author}{\bibfnamefont{G.}~\bibnamefont{Platero}},
  \bibinfo{journal}{Phys. Rev. Lett.} \textbf{\bibinfo{volume}{110}},
  \bibinfo{pages}{200403} (\bibinfo{year}{2013}).

\bibitem[{\citenamefont{Fruchart}(2016)}]{Fruchart2016}
\bibinfo{author}{\bibfnamefont{M.}~\bibnamefont{Fruchart}},
  \bibinfo{journal}{Phys. Rev. B} \textbf{\bibinfo{volume}{93}},
  \bibinfo{pages}{115429} (\bibinfo{year}{2016}).

\bibitem[{\citenamefont{Nathan and Rudner}(2015)}]{Frederik2015}
\bibinfo{author}{\bibfnamefont{F.}~\bibnamefont{Nathan}} \bibnamefont{and}
  \bibinfo{author}{\bibfnamefont{M.~S.} \bibnamefont{Rudner}},
  \bibinfo{journal}{New Journal of Physics} \textbf{\bibinfo{volume}{17}},
  \bibinfo{pages}{125014} (\bibinfo{year}{2015}).

\bibitem[{\citenamefont{Roy and Harper}(2017)}]{Roy2017}
\bibinfo{author}{\bibfnamefont{R.}~\bibnamefont{Roy}} \bibnamefont{and}
  \bibinfo{author}{\bibfnamefont{F.}~\bibnamefont{Harper}},
  \bibinfo{journal}{Phys. Rev. B} \textbf{\bibinfo{volume}{96}},
  \bibinfo{pages}{155118} (\bibinfo{year}{2017}).

\bibitem[{\citenamefont{Iadecola et~al.}(2015)\citenamefont{Iadecola, Santos,
  and Chamon}}]{Iadecola2015}
\bibinfo{author}{\bibfnamefont{T.}~\bibnamefont{Iadecola}},
  \bibinfo{author}{\bibfnamefont{L.~H.} \bibnamefont{Santos}},
  \bibnamefont{and} \bibinfo{author}{\bibfnamefont{C.}~\bibnamefont{Chamon}},
  \bibinfo{journal}{Phys. Rev. B} \textbf{\bibinfo{volume}{92}},
  \bibinfo{pages}{125107} (\bibinfo{year}{2015}).

\bibitem[{\citenamefont{Carpentier et~al.}(2015)\citenamefont{Carpentier,
  Delplace, Fruchart, and Gawedzki}}]{Carpentier2015}
\bibinfo{author}{\bibfnamefont{D.}~\bibnamefont{Carpentier}},
  \bibinfo{author}{\bibfnamefont{P.}~\bibnamefont{Delplace}},
  \bibinfo{author}{\bibfnamefont{M.}~\bibnamefont{Fruchart}}, \bibnamefont{and}
  \bibinfo{author}{\bibfnamefont{K.}~\bibnamefont{Gawedzki}},
  \bibinfo{journal}{Phys. Rev. Lett.} \textbf{\bibinfo{volume}{114}},
  \bibinfo{pages}{106806} (\bibinfo{year}{2015}).

\bibitem[{\citenamefont{Ho and Gong}(2014)}]{Ho2014}
\bibinfo{author}{\bibfnamefont{D.~Y.~H.} \bibnamefont{Ho}} \bibnamefont{and}
  \bibinfo{author}{\bibfnamefont{J.}~\bibnamefont{Gong}},
  \bibinfo{journal}{Phys. Rev. B} \textbf{\bibinfo{volume}{90}},
  \bibinfo{pages}{195419} (\bibinfo{year}{2014}).

\bibitem[{\citenamefont{Bomantara et~al.}(2016)\citenamefont{Bomantara,
  Raghava, Zhou, and Gong}}]{Bomantara2016}
\bibinfo{author}{\bibfnamefont{R.~W.} \bibnamefont{Bomantara}},
  \bibinfo{author}{\bibfnamefont{G.~N.} \bibnamefont{Raghava}},
  \bibinfo{author}{\bibfnamefont{L.}~\bibnamefont{Zhou}}, \bibnamefont{and}
  \bibinfo{author}{\bibfnamefont{J.}~\bibnamefont{Gong}},
  \bibinfo{journal}{Phys. Rev. E} \textbf{\bibinfo{volume}{93}},
  \bibinfo{pages}{022209} (\bibinfo{year}{2016}).

\bibitem[{\citenamefont{Jiang et~al.}(2011)\citenamefont{Jiang, Kitagawa,
  Alicea, Akhmerov, Pekker, Refael, Cirac, Demler, Lukin, and Zoller}}]{Jiang}
\bibinfo{author}{\bibfnamefont{L.}~\bibnamefont{Jiang}},
  \bibinfo{author}{\bibfnamefont{T.}~\bibnamefont{Kitagawa}},
  \bibinfo{author}{\bibfnamefont{J.}~\bibnamefont{Alicea}},
  \bibinfo{author}{\bibfnamefont{A.~R.} \bibnamefont{Akhmerov}},
  \bibinfo{author}{\bibfnamefont{D.}~\bibnamefont{Pekker}},
  \bibinfo{author}{\bibfnamefont{G.}~\bibnamefont{Refael}},
  \bibinfo{author}{\bibfnamefont{J.~I.} \bibnamefont{Cirac}},
  \bibinfo{author}{\bibfnamefont{E.}~\bibnamefont{Demler}},
  \bibinfo{author}{\bibfnamefont{M.~D.} \bibnamefont{Lukin}}, \bibnamefont{and}
  \bibinfo{author}{\bibfnamefont{P.}~\bibnamefont{Zoller}},
  \bibinfo{journal}{Phys. Rev. Lett.} \textbf{\bibinfo{volume}{106}},
  \bibinfo{pages}{220402} (\bibinfo{year}{2011}).

\bibitem[{\citenamefont{Thakurathi et~al.}(2013)\citenamefont{Thakurathi,
  Patel, Sen, and Dutta}}]{Thakurathi2013}
\bibinfo{author}{\bibfnamefont{M.}~\bibnamefont{Thakurathi}},
  \bibinfo{author}{\bibfnamefont{A.~A.} \bibnamefont{Patel}},
  \bibinfo{author}{\bibfnamefont{D.}~\bibnamefont{Sen}}, \bibnamefont{and}
  \bibinfo{author}{\bibfnamefont{A.}~\bibnamefont{Dutta}},
  \bibinfo{journal}{Phys. Rev. B} \textbf{\bibinfo{volume}{88}},
  \bibinfo{pages}{155133} (\bibinfo{year}{2013}).

\bibitem[{\citenamefont{Kundu and Seradjeh}(2013)}]{Kundu2013}
\bibinfo{author}{\bibfnamefont{A.}~\bibnamefont{Kundu}} \bibnamefont{and}
  \bibinfo{author}{\bibfnamefont{B.}~\bibnamefont{Seradjeh}},
  \bibinfo{journal}{Phys. Rev. Lett.} \textbf{\bibinfo{volume}{111}},
  \bibinfo{pages}{136402} (\bibinfo{year}{2013}).

\bibitem[{\citenamefont{Wang et~al.}(2017)\citenamefont{Wang, Chen, Bomantara,
  Gong, and Xing}}]{Wang2017}
\bibinfo{author}{\bibfnamefont{H.-Q.} \bibnamefont{Wang}},
  \bibinfo{author}{\bibfnamefont{M.~N.} \bibnamefont{Chen}},
  \bibinfo{author}{\bibfnamefont{R.~W.} \bibnamefont{Bomantara}},
  \bibinfo{author}{\bibfnamefont{J.}~\bibnamefont{Gong}}, \bibnamefont{and}
  \bibinfo{author}{\bibfnamefont{D.~Y.} \bibnamefont{Xing}},
  \bibinfo{journal}{Phys. Rev. B} \textbf{\bibinfo{volume}{95}},
  \bibinfo{pages}{075136} (\bibinfo{year}{2017}).

\bibitem[{\citenamefont{Wang et~al.}(2013)\citenamefont{Wang, Steinberg,
  Jarillo-Herrero, and Gedik}}]{Wang2013Observation}
\bibinfo{author}{\bibfnamefont{Y.~H.} \bibnamefont{Wang}},
  \bibinfo{author}{\bibfnamefont{H.}~\bibnamefont{Steinberg}},
  \bibinfo{author}{\bibfnamefont{P.}~\bibnamefont{Jarillo-Herrero}},
  \bibnamefont{and} \bibinfo{author}{\bibfnamefont{N.}~\bibnamefont{Gedik}},
  \bibinfo{journal}{Science} \textbf{\bibinfo{volume}{342}},
  \bibinfo{pages}{453} (\bibinfo{year}{2013}).

\bibitem[{\citenamefont{Lindner et~al.}(2010)\citenamefont{Lindner, Refael, and
  Galitski}}]{Lindner2010Floquet}
\bibinfo{author}{\bibfnamefont{N.}~\bibnamefont{Lindner}},
  \bibinfo{author}{\bibfnamefont{G.}~\bibnamefont{Refael}}, \bibnamefont{and}
  \bibinfo{author}{\bibfnamefont{V.}~\bibnamefont{Galitski}},
  \bibinfo{journal}{Nat. Phys.} \textbf{\bibinfo{volume}{7}},
  \bibinfo{pages}{490} (\bibinfo{year}{2010}).

\bibitem[{\citenamefont{Kraus et~al.}(2012)\citenamefont{Kraus, Lahini, Ringel,
  Verbin, and Zilberberg}}]{Kraus2012}
\bibinfo{author}{\bibfnamefont{Y.~E.} \bibnamefont{Kraus}},
  \bibinfo{author}{\bibfnamefont{Y.}~\bibnamefont{Lahini}},
  \bibinfo{author}{\bibfnamefont{Z.}~\bibnamefont{Ringel}},
  \bibinfo{author}{\bibfnamefont{M.}~\bibnamefont{Verbin}}, \bibnamefont{and}
  \bibinfo{author}{\bibfnamefont{O.}~\bibnamefont{Zilberberg}},
  \bibinfo{journal}{Phys. Rev. Lett.} \textbf{\bibinfo{volume}{109}},
  \bibinfo{pages}{106402} (\bibinfo{year}{2012}).

\bibitem[{\citenamefont{Khanikaev et~al.}(2013)\citenamefont{Khanikaev,
  Mousavi, Tse, Kargarian, Macdonald, and Shvets}}]{Khanikaev2013Photonic}
\bibinfo{author}{\bibfnamefont{A.~B.} \bibnamefont{Khanikaev}},
  \bibinfo{author}{\bibfnamefont{S.~H.} \bibnamefont{Mousavi}},
  \bibinfo{author}{\bibfnamefont{W.~K.} \bibnamefont{Tse}},
  \bibinfo{author}{\bibfnamefont{M.}~\bibnamefont{Kargarian}},
  \bibinfo{author}{\bibfnamefont{A.~H.} \bibnamefont{Macdonald}},
  \bibnamefont{and} \bibinfo{author}{\bibfnamefont{G.}~\bibnamefont{Shvets}},
  \bibinfo{journal}{Nat. Mater.} \textbf{\bibinfo{volume}{12}},
  \bibinfo{pages}{233} (\bibinfo{year}{2013}).

\bibitem[{\citenamefont{Khanikaev and Shvets}(2017)}]{khanikaev2017two}
\bibinfo{author}{\bibfnamefont{A.~B.} \bibnamefont{Khanikaev}}
  \bibnamefont{and} \bibinfo{author}{\bibfnamefont{G.}~\bibnamefont{Shvets}},
  \bibinfo{journal}{Nature Photonics} \textbf{\bibinfo{volume}{11}},
  \bibinfo{pages}{763} (\bibinfo{year}{2017}).

\bibitem[{\citenamefont{Lu et~al.}(2014)\citenamefont{Lu, Joannopoulos, and
  Solja?i?}}]{Lu2014Topological}
\bibinfo{author}{\bibfnamefont{L.}~\bibnamefont{Lu}},
  \bibinfo{author}{\bibfnamefont{J.~D.} \bibnamefont{Joannopoulos}},
  \bibnamefont{and} \bibinfo{author}{\bibfnamefont{M.}~\bibnamefont{Solja?i?}},
  \bibinfo{journal}{Nature Photonics} \textbf{\bibinfo{volume}{8}},
  \bibinfo{pages}{821} (\bibinfo{year}{2014}).

\bibitem[{\citenamefont{Rechtsman et~al.}(2013)\citenamefont{Rechtsman, Zeuner,
  Plotnik, Lumer, Podolsky, Dreisow, Nolte, Segev, and
  Szameit}}]{Rechtsman2013Photonic}
\bibinfo{author}{\bibfnamefont{M.~C.} \bibnamefont{Rechtsman}},
  \bibinfo{author}{\bibfnamefont{J.~M.} \bibnamefont{Zeuner}},
  \bibinfo{author}{\bibfnamefont{Y.}~\bibnamefont{Plotnik}},
  \bibinfo{author}{\bibfnamefont{Y.}~\bibnamefont{Lumer}},
  \bibinfo{author}{\bibfnamefont{D.}~\bibnamefont{Podolsky}},
  \bibinfo{author}{\bibfnamefont{F.}~\bibnamefont{Dreisow}},
  \bibinfo{author}{\bibfnamefont{S.}~\bibnamefont{Nolte}},
  \bibinfo{author}{\bibfnamefont{M.}~\bibnamefont{Segev}}, \bibnamefont{and}
  \bibinfo{author}{\bibfnamefont{A.}~\bibnamefont{Szameit}},
  \bibinfo{journal}{Nature} \textbf{\bibinfo{volume}{496}},
  \bibinfo{pages}{196} (\bibinfo{year}{2013}).

\bibitem[{\citenamefont{Maczewsky et~al.}(2017)\citenamefont{Maczewsky, Zeuner,
  Nolte, and Szameit}}]{maczewsky2017observation}
\bibinfo{author}{\bibfnamefont{L.~J.} \bibnamefont{Maczewsky}},
  \bibinfo{author}{\bibfnamefont{J.~M.} \bibnamefont{Zeuner}},
  \bibinfo{author}{\bibfnamefont{S.}~\bibnamefont{Nolte}}, \bibnamefont{and}
  \bibinfo{author}{\bibfnamefont{A.}~\bibnamefont{Szameit}},
  \bibinfo{journal}{Nat. commun.} \textbf{\bibinfo{volume}{8}},
  \bibinfo{pages}{13756} (\bibinfo{year}{2017}).

\bibitem[{\citenamefont{Su et~al.}(1979)\citenamefont{Su, Schrieffer, and
  Heeger}}]{Su1979}
\bibinfo{author}{\bibfnamefont{W.~P.} \bibnamefont{Su}},
  \bibinfo{author}{\bibfnamefont{J.~R.} \bibnamefont{Schrieffer}},
  \bibnamefont{and} \bibinfo{author}{\bibfnamefont{A.~J.}
  \bibnamefont{Heeger}}, \bibinfo{journal}{Phys. Rev. Lett.}
  \textbf{\bibinfo{volume}{42}}, \bibinfo{pages}{1698} (\bibinfo{year}{1979}).

\bibitem[{\citenamefont{Rice and Mele}(1982)}]{Rice}
\bibinfo{author}{\bibfnamefont{M.~J.} \bibnamefont{Rice}} \bibnamefont{and}
  \bibinfo{author}{\bibfnamefont{E.~J.} \bibnamefont{Mele}},
  \bibinfo{journal}{Phys. Rev. Lett.} \textbf{\bibinfo{volume}{49}},
  \bibinfo{pages}{1455} (\bibinfo{year}{1982}).

\bibitem[{\citenamefont{Shen}(2017)}]{Shen2017Topological}
\bibinfo{author}{\bibfnamefont{S.~Q.} \bibnamefont{Shen}},
  \emph{\bibinfo{title}{Topological insulators : Dirac equation in condensed
  matters}} (\bibinfo{publisher}{Springer}, \bibinfo{year}{2017}).

\bibitem[{\citenamefont{Zhou et~al.}(2015)\citenamefont{Zhou, Tan, and
  Gong}}]{Zhou2015}
\bibinfo{author}{\bibfnamefont{L.}~\bibnamefont{Zhou}},
  \bibinfo{author}{\bibfnamefont{D.~Y.} \bibnamefont{Tan}}, \bibnamefont{and}
  \bibinfo{author}{\bibfnamefont{J.}~\bibnamefont{Gong}},
  \bibinfo{journal}{Phys. Rev. B} \textbf{\bibinfo{volume}{92}},
  \bibinfo{pages}{245409} (\bibinfo{year}{2015}).

\bibitem[{\citenamefont{Wang et~al.}(2015)\citenamefont{Wang, Zhou, and
  Gong}}]{Wang2015Interband}
\bibinfo{author}{\bibfnamefont{H.}~\bibnamefont{Wang}},
  \bibinfo{author}{\bibfnamefont{L.}~\bibnamefont{Zhou}}, \bibnamefont{and}
  \bibinfo{author}{\bibfnamefont{J.}~\bibnamefont{Gong}},
  \bibinfo{journal}{Phys. Rev. B} \textbf{\bibinfo{volume}{91}},
  \bibinfo{pages}{085420} (\bibinfo{year}{2015}).

\bibitem[{\citenamefont{Ho and Gong}(2012)}]{Ho2012}
\bibinfo{author}{\bibfnamefont{D.~Y.~H.} \bibnamefont{Ho}} \bibnamefont{and}
  \bibinfo{author}{\bibfnamefont{J.}~\bibnamefont{Gong}},
  \bibinfo{journal}{Phys. Rev. Lett.} \textbf{\bibinfo{volume}{109}},
  \bibinfo{pages}{010601} (\bibinfo{year}{2012}).

\end{thebibliography}
\end{document}